\documentclass[authoryear,preprint,10pt,review]{elsarticle}
\usepackage{amssymb,amsmath,gensymb,multirow}
\usepackage{booktabs}
\usepackage{lineno}
\usepackage{float}
\usepackage{color}
\journal{arXiv } 
\begin{document}
\begin{frontmatter}
\title{Back to the future: a simplified and intuitive derivation of the Lotka-Euler equation}

\author[UFG]{Carlos Hernandez-Suarez\corref{cor1}}
 \ead{cmh1@cornell.edu}

 \address[UC]{Instituto de Ciencias Tecnolog\'ia e Innovaci\'on, Universidad Francisco Gavidia, El Progreso St., No. 2748, Colonia Flor Blanca, San Salvador, El Salvador}
  
 \cortext[cor1]{Corresponding author}

\begin{abstract}

The Lotka-Euler equation is a mathematical expression used to study population dynamics and growth, particularly in the context of demography and ecology. The growth rate $\lambda$ is the speed at which an individual produce their offspring. It is essentially a birth process, and here it is shown that by reversing the process to a death process, in which individuals die at a rate $\lambda^{-1}$, the derivation of the Lotka-Euler equation becomes more intuitive and direct, both in discrete and continuous time.

\end{abstract}

\begin{keyword}
Lotka-Euler \sep Growth rate \sep Basic reproductive number
\end{keyword}

\end{frontmatter}

\section{Introduction}

The Lotka-Euler equation serve as valuable tools for analyzing and modeling population growth rates, age distribution of individuals, and reproductive patterns within a population. This equation is predominantly employed in the study of stable age populations, where age-specific birth and death rates remain constant over time. The Lotka-Euler equation was initially formulated for continuous time by \cite{lotka1907relation}\cite[ see also][]{lotka1907art,sharpe1911problem}, while its discrete-time counterpart was developed by Cole \cite{cole1954population}. Generally, the growth rate is found by numerically solving the Lotka-Euler equation. Recently \cite{Hernandez-Suarez2023.04.12.536246} derived exact confidence intervals for this parameter.

\section{The Lotka-Euler equation}

First we present the Lotka-Euler equation reproduced from \cite{wallinga2007generation}, that focus on female individuals, under the assumption that there is plenty of males for reproductive purposes. We assume that the population grows exponetially at a fixed rate $r$ and that the population is at a stable age distribution.

The Lotka-Euler equation can be understood as an integration of two concepts: first, by summing the number of children born to mothers of all ages at a specific time, we can determine the total number of births for that time. The number of births to mothers of age $a$ at time $t$ is equivalent to the number of births at time $t-a$ (including mothers who did not survive) multiplied by the expected number of offspring per year for mothers of age $a$. By adding up these births across all possible ages of mothers, we can calculate the total number of births in un it oif time $t$t:

$$
b(t) = \int_0^\infty b(t-a) m(a) \ \text{d}a
$$

In this equation, $b(t)$ denotes the population's birth rate at time $t$, while $n(a)$ is the rate at which a mother at age $a$ produces female offspring. Given that the population is experiencing exponential growth with a consistent age distribution, the number of births at any particular time (for example, $t$) is equivalent to the number of births a time units ago, multiplied by the population's exponential growth since that time:

$$
b(t) = b(t-a) e^{r a}
$$

By integrating these two equations, we derive an expression with $b(t)$ on both sides:

$$
b(t) = \int_0^\infty b(t) e^{-r a} m(a) \ \text{d}a
$$

This equation can be intuitively interpreted as follows: the sum of all past births, multiplied by the number of offspring for individuals born at each time in the past, must equal the current number of births. In demographic terms, the composite parameter $n(a)$ is more commonly recognized as the product of the survivorship and fecundity functions, denoted by $n(a)=l(a)m(a)$. By employing this more familiar parameterization and eliminating $b(t)$ from both sides of the equation, we arrive at the Lotka-Euler equation:

\begin{equation*}
1= \int_0^\infty e^{-r a} \ l(a) m(a) \ \text{d}a. 
\end{equation*}
If we integrate $n(a)$ over the whole lifespan, we obtain the total number of female offspring produced by a mother over her lifespan, known as $R_0$:
\begin{equation*}
\int_0^\infty n(a) \ \text{d}a = R_0. 
\end{equation*}
The rate $n(a)$ can be normalized to a distribution$ g(a)$, which is the distribution of age at child bearing:
$$
g(a) =\frac{n(a)}{R_0}
$$
thus we finally arrive to:
\begin{equation}
\int_0^\infty R_0 \ g(a) e^{-r a} \text{d}a=1 \label{eq:LE}
\end{equation}

Cole  (\ref{eq:Cole}) derived a discrete time equivalent to the Lotka-Euler equation, that we do not reconstruct here, arriving to:

\begin{equation}
\sum_{x=1}^\infty e^{-r x} \ l(x) m(x)=1 \label{eq:Cole}
\end{equation}
In what follows we present alternative derivations for both equations,

\section{Alternative construction of the Lotka-Euler equation}
\subsection{Discrete time}
First we derive the equivalent to equation (\ref{eq:LE}) for discrete time. In this setting, events occur in time intervals represented by $1, 2, 3, \ldots$. At this point, it is unnecessary to determine if events occur at the beginning or the end of an interval. The focal point is their occurrence within a specific interval.

First assume $\lambda >1$, which allows to derive an intuitive result. Suppose that the population experiences growth at a rate of $\lambda$ per unit of time. Starting with a single individual, the population size after $t$ units of time is expressed as:
\begin{equation}
\lambda^t = N \label{eq:eq1}
\end{equation}
Given known values of $t$ and $N$, the growth rate can with estimated by $\lambda = N^{1/t}$. However, equation (\ref{eq:eq1}) also implies that
\begin{equation}
N (1/\lambda)^t = 1 \label{eq:eq2}
\end{equation}
which admits the following interpretation: if a population starts with $N$ individuals, who die at a rate of $\lambda^{-1}$, the population will reduce to a size of 1 after $t$ units of time. Clearly, despite being now a death process rather than a birth process, the solution for $\lambda$ in equation (\ref{eq:eq2}) remains identical.

The growth rate represents the rate at which individuals produce their offspring of average size $R_0$, also known as the \textit{basic reproductive number}. Let $g(x)$ denote the probability mass function of the age of mothers at the time of bearing. Consider first the case in which all mothers give birth to their entire progeny at age $x$. The growth rate, as deduced from equation (\ref{eq:eq2}) and the rationale supporting it, is the solution of $\lambda$ such that:

\begin{equation}
\frac{R_0}{\lambda^x} = 1, 
\end{equation}
if only a fraction $g(x)$ is born when mothers are of age $x$, then:
\begin{equation}
\frac{g(x) R_0}{\lambda^x} = g(x). 
\end{equation}
which implies that the fraction that is born of mothers of age $x$ produce a fraction $g(x)$ of the progeny. Summing over $x$ yields;
\begin{equation}
\sum_x \frac{g(x) R_0}{\lambda^x} = \sum_x g(x). 
\end{equation}
that is:
\begin{equation}
\sum_x \frac{g(x) R_0}{\lambda^x} = 1. \label{eq:eq5}
\end{equation}
Since $g(x) = l(x) \ m(x)/R_0$, equation (\ref{eq:eq5}) is identical the Lotka-Euler equivalent expression for discrete time  (\ref{eq:Cole}), derived in \cite{cole1954population}, using $\lambda$ instead of $e^r$. 

\subsection{Continuous time}
For the equivalent continuous-time expression, the age at which offspring are produced, $g(t)$, must be a continuous function exhibiting properties similar to those of a probability density function. Furthermore, the instantaneous growth rate, $r$, is defined as the value satisfying

\begin{equation}
\lim_{n \rightarrow \infty} (1+r/n)^n = \lambda \label{eq:eq6}
\end{equation}
which results in $r=\log(\lambda)$. The continuous-time counterpart to equation (\ref{eq:eq5}) is thus:
\begin{equation}
\int_0^\infty R_0 \ g(t) e^{-r t} \text{d}t=1 \label{eq:eq7}
\end{equation}
which is identical to (\ref{eq:LE}).

If $\lambda < 1$ then $N > N R_0$ the population shrinks,  but all equations (\ref{eq:eq1} -\ref{eq:eq7}) still apply.


\section{Bibliography}
\begin{thebibliography}{6}
\providecommand{\natexlab}[1]{#1}
\providecommand{\url}[1]{\texttt{#1}}
\providecommand{\urlprefix}{URL }
\expandafter\ifx\csname urlstyle\endcsname\relax
  \providecommand{\doi}[1]{doi:\discretionary{}{}{}#1}\else
  \providecommand{\doi}{doi:\discretionary{}{}{}\begingroup
  \urlstyle{rm}\Url}\fi

\bibitem[{Cole(1954)}]{cole1954population}
Cole, L.C. (1954).
\newblock The population consequences of life history phenomena.
\newblock \textit{The Quarterly review of biology} 29(2): 103--137.

\bibitem[{Hernandez-Suarez and
  Rabinovich(2023)}]{Hernandez-Suarez2023.04.12.536246}
Hernandez-Suarez, C. and Rabinovich, J. (2023).
\newblock Exact confidence intervals for population growth rate, longevity and
  generation time.
\newblock \textit{bioRxiv} \doi{10.1101/2023.04.12.536246}.

\bibitem[{Lotka(1907{\natexlab{a}})}]{lotka1907relation}
Lotka, A.J. (1907{\natexlab{a}}).
\newblock Relation between birth rates and death rates.
\newblock \textit{Science} 26(653): 21--22.

\bibitem[{Lotka(1907{\natexlab{b}})}]{lotka1907art}
Lotka, A.J. (1907{\natexlab{b}}).
\newblock Studies on the mode of growth of material aggregates.
\newblock \textit{American Journal of Science (1880-1910)} 24(141): 199.

\bibitem[{Sharpe and Lotka(1911)}]{sharpe1911problem}
Sharpe, F.R. and Lotka, A.J. (1911).
\newblock L. a problem in age-distribution.
\newblock \textit{The London, Edinburgh, and Dublin Philosophical Magazine and
  Journal of Science} 21(124): 435--438.

\bibitem[{Wallinga and Lipsitch(2007)}]{wallinga2007generation}
Wallinga, J. and Lipsitch, M. (2007).
\newblock How generation intervals shape the relationship between growth rates
  and reproductive numbers.
\newblock \textit{Proceedings of the Royal Society B: Biological Sciences}
  274(1609): 599--604.
\newblock \doi{http://doi.org/10.1098/rspb.2006.3754}.

\end{thebibliography}
\end{document}